\documentclass[a4paper,fleqn,usenatbib]{mnras}

\usepackage[T1]{fontenc}
\usepackage{ae,aecompl}

\usepackage{graphicx}   % Including figure files
\usepackage{amsmath}    % Advanced maths commands
\usepackage{amssymb}    % Extra maths symbols

\newcommand{\etal}{et al.}
\def\lesssim{\mathrel{\hbox{\rlap{\hbox{\lower4pt\hbox{$\sim$}}}\hbox{$<$}}}}
\def\gtrsim{\mathrel{\hbox{\rlap{\hbox{\lower4pt\hbox{$\sim$}}}\hbox{$>$}}}}
\def\apj{ApJ}
\def\apjs{ApJS}
\def\aj{AJ}
\def\aap{A\&\hskip-1pt A}

\def\mnras{MNRAS}

\def\araa{ARA\&\hskip-1pt A}

\def\jkas{JKAS}
\title[SFHs of dSphs and dEs]{Star Formation Histories of Dwarf Spheroidal and
Dwarf Elliptical Galaxies in the Local Universe}
\author[Seo \& Ann]{
Mira Seo$^1$ and Hong Bae Ann$^{1}$\thanks{E-mail: hbann@pusan.ac.kr}
\\
$^{1}$Department of Earth Sciences, Pusan National University, Busan, 609-735, Korea\\
}
\date{Accepted XXX. Received YYY; in original form ZZZ}

\pubyear{2019}

\begin{document}
\label{firstpage}
\pagerange{\pageref{firstpage}--\pageref{lastpage}}
\maketitle

%%\title{Star Formation Histories of dSph  and dE Galaxies in the Local Universe}

%% you desire. In v6.2 you can include a footnote in the title.

\begin{abstract}

We present the star formation histories (SFHs) of early-type dwarf galaxies, dSphs and dEs, in the local universe within z = 0.01.
The SFHs of early-type dwarf galaxies are characterized by
pre-enriched, metal-poor old stellar populations, absence of moderately old
stars that have ages of a few Gyr.
There are some differences in the SFHs of dSphs and dEs. In particular,
dSphs formed old ($\gtrsim10$ Gyr old) metal-poor stars $\sim2$ times
more than dEs.
The effects of reionization and feedback from supernova explosions are
thought to be strong enough to remove the gas left, which prevent
moderately old stellar populations in dSphs. In contrast,
the ejected gas are not completely removed from dEs and fall back
to ignite burst of star formation at a few Gyr after the first period of violent bursts of star formation, showing a suppression of star formation at lookback time $\approx 9.6$ Gyr. The second peak of star formation at lookback time $\approx 4.5$ Gyr in dEs produce moderately old stellar populations. Distinction between dSphs and dEs is useful to examine the SFHS of the early-type dwarfs since the cumulative SFHs are most closely related to their morphology. The stellar mass plays an important role in the SFHs of the early-type dwarfs as a driver of star formation, especially in galaxies with primordial origin.  

\end{abstract}


\section{Introduction}

Dwarf galaxies are most abundant in the local universe. They are thought to
be building blocks of more massive galaxies in the framework of $\Lambda$ cold
dark matter ($\Lambda$CDM) cosmology \citep{blu85, nav97}. Some dwarf galaxies
survive to the present time as isolated galaxies or satellites of larger galaxies. Among dwarf galaxies, early-type dwarf galaxies such as dwarf spheroidal galaxies (dSphs) and dwarf elliptical galaxies (dEs) are of special interest because they could be primordial objects that evolved to the present form. It is also likely that they are transformed ones from late-type galaxies if they are observed in group/cluster environment where a variety of mechanisms that transform late-type galaxies to early-type dwarfs are operable \citep[e.g.,][references therein]{fat18}.

Some early-type dwarf galaxies that have primordial origin are thought to be fossils of primordial galaxies whose star formation had quenched before or during re-ionization \citep{gut22}. The majority of stars in these galaxies have ages older than 10 Gyr with metallicities lower than [Fe/H]$\approx-2$ since re-ionization is thought to occur at $Z\approx6$ \citep{bec01}. Quenching of star formation was driven by the mechanical feedback from supernova explosions followed by the bursts of star formation at an early epoch of cosmic evolution. Mechanical energy from the supernova explosions blows out the gas from the primordial galaxies which have shallow potentials due to small masses \citep{vad86, dek86, sal08}. Examples of such fossil objects are the Sextans dwarf spheroidal galaxy in the LG \citep{bet18}. The ultra faint dwarf spheroidal (UDF) Segue 1 is also supposed to be fossil object \citep{fre14, web16}.

The dEs comprise the most abundant population of galaxy clusters such as the Virgo cluster \citep{bin85} while the dSphs are mostly found in the Local Group (LG). Whether they are morphologically distinct or not, they are mostly satellites of bright giant galaxies \citep{mak15, ann17}.
In the Local Group, most satellites of the Milky Way (MW) and Andromeda (M31), located within $\sim300$ kpc from their host galaxies, are dSphs \citep{mcc12}. They are thought to be accreted into the host galaxies earlier than $\sim6$ Gyr ago \citep{sou21, bar23} but \citet{ham23} argued that most dwarfs of MW are newcomers based on energy arguments. Some of them could
be primordial objects but the majority of them are thought to be transformed
from gas-rich dwarf irregulars which lost their cold gas by tidal stripping \citep{moo96, moo98, may01a, may01b, kra04, pau14, kaz17, fat18} and/or ram pressure stripping \citep{gg72, aba99, koo18}. The spiral arm features revealed in the unsharp masked images of a number of dEs in the Virgo clusters \citep{jer00, lis06} support the transformed late-type galaxies as their origin. The gas-rich late-type galaxies can be deprived of their gas by ram pressure stripping  \citep{gg72} and starvation \citep{lar80}.

Besides the favorable environment for morphology transformation for early-type dwarfs, there are some observations that support the non-primordial origin of dEs. They are the embedded spirals observed in some dEs \citep{jer00, bara02, gra03, rij03, lis06} and fast rotations which imply rotationally supported systems \citep{ped02, sim02, geh03, zee04, rij05, chi09, geh10, tol15, pen16} together with the blue-cored dEs, which implies presence of cold gas at the central regions of a significant number of dEs in the Virgo Cluster \citep{lis07, ham19}.  The presence of blue-cored dEs are also observed in the Fornax Cluster and Coma Cluster with much larger fractions \citep{ham19}.  Actually, there are some HI gas-rich dEs in the Virgo Cluster \citep{con03, hal12} and in the Fornax Cluster \citep{buy05}.  On the other hand, there has been no report on the embedded spirals in dSphs. Most dSphs are dispersion-supported systems except for some dSphs which are supposed to be formed by mergers \citep{lok14}. But the paucity of rotation-supported dSphs does not necessarily admit that they are not transformed from late-type galaxies because the dwarf irregulars seem to be dispersion-supported \citep{whe17}.

A non-negligible fraction of dEs and dSphs are isolated early-type dwarf galaxies of which origins could be primordial objects because they have no chance to interact with other galaxies. A recent CMD analysis for the star formation history (SFH) of an early-type dwarf galaxy Sextans \citep{bet18} shows that the majority of the stellar component formed before the end of the reionization which is assumed to be fixed at $\sim12.77$ Gyr \citep{bec01}. Existence of early-type dwarf galaxies whose cold gas was completely exhausted in the early phase of galaxy evolution suggests different gas-loss mechanisms. Examples of the proposed gas-loss mechanisms are photoevaporation \citep{bar99,  gne00, sha04}, supernova (SN) feedback \citep{dek86, fer00, mar06, bov09}, interaction with gaseous filaments \citep{ben13}. Dissolving star cluster scenario \citep{ass13} is another way to explain the formation of isolated dSph galaxies.

There has been no consensus on the morphological classification of dwarf galaxies. Simple classification scheme is to divide them as early-type dwarfs and late-type dwarfs. Aside from the late-type dwarfs which are designated by dIrrs or dIs, early-type dwarfs are frequently divided into two sub-types of galaxies, dEs and dS0s, where the
former include their faint cousins dSphs \citep{san84}. Most studies \citep[eg.,][]{bin85, lis06} used the dE/dS0 division for early-type dwarf galaxies following \citet{san84} although the terminology dwarf spheroidals (dSphs) have been used for the satellite galaxies of the MW and M31. \citet{kor12} also neglect the sub-type dSph in proposing a new scheme for morphological classification which extends the parallel sequence of \citet{ber76}. They used Sph rather than dE to represent the early-type dwarfs which include dE and dS0 galaxies in \citet{bin85} and considered that dSphs are merely a smaller version of dEs. \citet{but15} also used the notation Sph to describe the dE/dS0 galaxies observed in the Spitzer Survey of Stellar Structure in Galaxies (S4G).  On the hand, the Updated  Nearby Galaxy Catalog \citep[UNGC]{kar13} employed a different approach to classifying galaxies in the Local Volume (LV) within 11 Mpc. They distinguished dSphs from dEs but they used  Sph rather than dSph to denote dSphs. They classify dwarf galaxies based on surface brightness and color. In their classification, dE and Sph galaxies are classified to have red colors but with different surface brightness: high surface brightness for dE and normal to extremely low surface brightness for Sph. For comparison, dS0 galaxies have normal surface brightness with red colors. Another example of dwarfs with high surface brightness is BCD but BCD galaxies have blue colors. The explicit use of dSph was made by \citet{ann15} to represent early-type dwarf galaxies similar to dE but shallower gradient in surface brightness distribution. The luminosity and colors of dSphs are a little bit less luminous and bluer than dEs but the majority of dSphs are brighter than M$_{r}$ = -13, which overlaps the luminosity range of dEs.

In contrast to the morphology classification grid proposed by \citet{gra19} where dwarf elliptical galaxies are treated as merely small versions of normal early-type galaxies, the early-type dwarfs, dSphs and dEs, can be distinguished from giant elliptical galaxies by their surface brightness distribution, which is better represented by
the  S\'{e}rsic profile \citep{ser68} with $n = 1-3$ \citep{geh02, seo22}.  There is a slight difference in the mean S\'{e}rsic index $n$ between dSphs and dEs in the sense that dSphs have smaller $n$ \citep{seo22} which leads to shallower surface brightness distribution of dSphs.
In contrast to the most previous studies \citep[eg.,][]{bin85, lis06, kor12}, we distinguish dSphs from dEs.
One of the reasons why we distinguish dSphs from the dEs is that dSph galaxies are mostly dispersion-supported systems \citep{wal09,sal12} while a considerable fraction of dE galaxies is supported by rotation \citep{geh10}.
Their origins might be different. The difference in luminosity of dEs and dSphs are apparent because there is no dEs fainter than $M_{B}\approx -13$, at least in the LG. However, In the luminosity ranges between $M_{R} = -13$ and $M_{R}=-16$,  a significant fraction of early-type dwarf galaxies is thought to be dSphs according to their morphology \citep{ann15}.

Since the morphology of a galaxy is an integral property that reflects the physical properties such as dynamical structure and stellar populations. The former property is revealed by the apparent shape and surface brightness distribution while the latter is reflected by the luminosity, color and metal abundance. Since the stellar population of a galaxy is determined by the SFH, galaxies with similar morphology may experience similar SFHs. Thus, we can expect a link between morphology and SFH of galaxies, although it is apparent that a considerable fraction of the satellite galaxies in the LG shows somewhat different SFHs for the same morphological type \citep{wei14a}. Since the SFHs of the 40 dwarf satellites of the LG were derived from the color-magnitude diagrams obtained from HST imaging, their SFHs are thought to be precise enough to reveal fine details of the evolution of these galaxies. The different SFHs of the dSphs in the LG might be due to differences in mass and environment. The mass range of the LG dSphs inferred from the luminosity range of $\sim 8$ mag is large enough to drive different SFHs in the early stage of evolution. Lower mass dwarfs are likely to be quenched earlier by the supernova feedback \citep{dek86, fer00, mar06, bov09}  than higher mass dwarfs. It is also not surprising that the star formation of the satellites at smaller distances from the host galaxy (MW or M31) is quenched earlier than those at a large distance by the environmental effects of their host halos \citep{bos08, pen12, zho20}.

Thus, it is necessary to analyze the stellar populations of early-type dwarfs to derive their SFHs. The most accurate method to analyze the stellar populations of a galaxy is to compare observed color-magnitude diagrams (CMDs) of resolved stars with synthetic CMDs \citep[references there in]{tol09}. But this method has been applied mostly to dwarf galaxies in the LG \citep{mon10a, mon10b, boe12, hid12, hid13, wei14a, gal15} because it requires deep imaging that reach the oldest main-sequence turn-off (MSTO).  Up to now, most distant galaxies of which CMDs are
deep enough to derive the oldest populations are limited within the local volume of 10 Mpc. Thus, it seems obligate to use integrated spectra of galaxies that are located outside the local volume of 10 Mpc. We used the population synthesis code STARLIGHT \citep{cidF05} which has been widely used for spectra observed in the Sloan Digital Sky Survey \citep[SDSS]{yor00}. Recently, there are a number of studies \citep[e.g.,][]{cid13}, applying STARLIGHT to explore stellar populations across galaxies using the spectra obtained from integral-field spectroscopic observations such as the Calar Alto Legacy Integral Field Area Survey \citep[CALIFA]{san12} and Mapping Nearby Galaxies at Apache Point Observatory \citep[MaNGA]{bun15}. Among others, the \citet{cid13} and \citet{rif21} are
examples of using STARLIGHT to the stellar population of galaxies in CALIFA and in MaNGA, respectively.

We are wondering that the morphological differences observed in the bright early-type dwarf galaxies are caused by the SFHs which are supposed to be closely related to their origins. The connection between the SFHs and the structural properties is also expected because the spatial distribution of stellar populations determines the structure of a galaxy. The flatter surface brightness distribution of dSphs which corresponds to smaller S\'{e}rsic index $n$ than dEs is thought to be caused by the shallow potential of the dSphs. We expect earlier quenching of star formation in dSphs than in dEs due to their mass difference. The difference in quenching epoch can be traced by the stellar populations.

Most previous studies of SFHs of dSphs and dEs using CMD analysis were confined to those in the LG \citep{her00, apa01, dol02, car02, lee09, mon10a, mon10b, boe12, hid12, hid13, bro14, wei14a, gal15, sav15, san16, mak17, ski17,  bet18, bet19, sav19, wei19, gal21,rus21, nav21} except for a few cases \citep[e.g.,][references therein]{wei11, cig19}. \citet{wei11} derived SFHs of 60 dwarf galaxies from the ACS Nearby Galaxy Survey Treasury (ANGST). They are located within $\sim4$ Mpc except for DDO 165 of which distance is 4.6 Mpc and masses in the range $M_{\ast} \sim10^{5-11} M_{\odot}$. This is because the derivation of SFHs requires high-quality CMDs that can be obtained for the galaxies of which constituent stars can be resolved.
Because of the low luminosity of dEs and dSphs, their SFHs derived from the color-magnitude diagrams are confined to those in the local volume, mostly those in the LG.

Population synthesis of the dwarf galaxies beyond the LG was performed for the 12 dSphs in Cen A group at a distance of  3.8 Mpc \citep{mul21} using integrated spectra obtained from the MUSE integral field spectrograph mounted at  UT4  of the  VLT  on  Cerro  Paranal,  Chile \citep{bac10}. The dSphs have old and metal-poor stellar populations and follow the stellar metallicity-luminosity relation defined by the dwarf galaxies in the Local Group.

Here, we report an analysis of the SFHs of the early-type dwarf galaxies listed in the catalogue of visually classified galaxies in the local ($z \lesssim 0.01$) universe \citep[CVCG]{ann15}. To derive the SFHs of the early-type dwarfs, we applied a population synthesis code STARLIGHT \citep{cidF05} to the optical spectra of two sub-types of the early-type dwarf galaxies (dSphs and dEs) observed in the Sloan Digital Sky Survey (SDSS; \citet{yor00}).

The present paper is organized as follows. In Section 2, selection of sample galaxies and observational data are described. The method and basic results of this study are presented in Section 3. The cumulative SFHs of the dSphs and dEs are described in section 4. Discussion of the SFHs of the early-type dwarfs is given in Section 5.
Summary and conclusions are given
in the last section.

\section{Sample Selection and Data}
\subsection{Selection of  dwarf spheroidals and dwarf ellipticals}

We selected the two sub-types of early-type dwarf galaxies, dSphs and dEs, from the galaxies in the CVCG which provides detailed morphological types of 5638 galaxies located in the local universe.
The galaxies listed in the CVCG are mostly taken from KIAS-VAGC \citep{cho10} which is a value-added
catalogue based on the Sloan Digital Sky Survey (SDSS) Data Release 7 (DR7). The CVCG supplements the original catalogue with galaxies brighter than $r=14.5$ whose redshifts were taken from literature \citep[references there in]{cho10}. The CVCG took about 1000 galaxies from the NASA Extragalactic Database (NED) which were not included in the KIAS-VAGC due to selection criteria adopted. Thus, the
 CVCG is nearly complete for galaxies brighter than $r=17.77$ for the regions surveyed by the SDSS. \citet{ann15} distinguished dSph and dE from other early-type morphology such as dS0 and  dE$_{bc}$, which are dwarf lenticular
galaxy and blue-cored dwarf elliptical galaxy, respectively.  We confined our sample galaxies to dSphs and dEs for population synthesis studies because morphological features such as disc and lens are observed in dS0s and young stellar populations are dominant in the cores of dE$_{bc}$ sub-types, which implies that they are not genuine primordial dwarfs although both of them are considered to be early-type dwarfs.

%%%%%%%%%%%%%%%%%%%%  Fig1 (sampleImage.png) %%%%%%%%%%%%%%%
\begin{figure}
\centering
\includegraphics[width=0.45\textwidth]{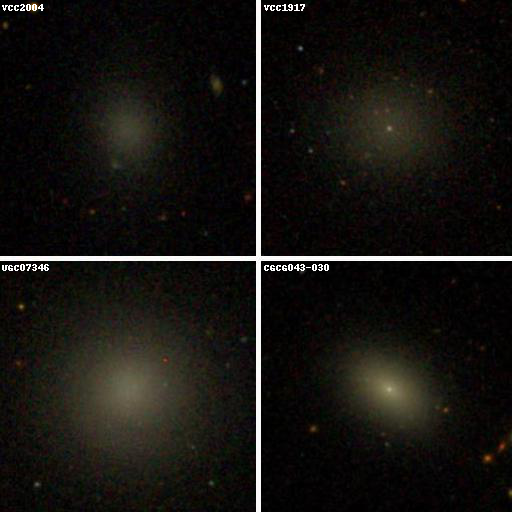}
\caption{SDSS color images of four early-type dwarfs. Galaxies in the upper row are dwarf spheroidal galaxies and those in the lower row are dwarf elliptical galaxies. Left panels for non-nucleated dwarfs and right panels for nucleated dwarfs. The box size in one dimension is $\sim100$ arcsec. North is up, and East to the left.}
%%\label{fig1}
\end{figure}
%\end{verbatim}%%%%%%%%%%%%%%%%%%%%%%%%%%%%%%%%%%%%%%%%%%%%%%%%%%%%%%%

Figure 1 shows the color images of four sample galaxies, the upper row for dSphs and the lower row for dEs. We provide nucleated and non-nucleated dwarfs for each sub-type.  Although the sub-types of early-type dwarfs are classified by the morphological characteristics \citep{ann15}, there are some difference in luminosity and colors of dSphs and dEs. The luminosity difference between dEs and dSphs is apparent because there is no dEs fainter than $M_{B}\approx -13$, at least in the LG. For the present sample of dSphs and dEs, on average, dSphs are $\sim1.5$ mag fainter than dEs. However, as shown in Figure 2, there is a range of luminosity,  $M_{r} =-13$ $\sim$ $M_{r}=-16$, where a significant fraction of dSphs and dEs are present. The majority of dSphs and dEs are located along the red sequence \citep{str01} but some deviate a lot from it. The large spread in the $u-r$ colors of dSphs seems to be due to large photometric errors in $u$-band magnitudes which occurred frequently in faint galaxies.

%%%%%%%%%%%%%%%%  Fig2  (urMr.eps) %%%%%%%%%%%%%
\begin{figure}
\centering
\includegraphics[width=0.43\textwidth]{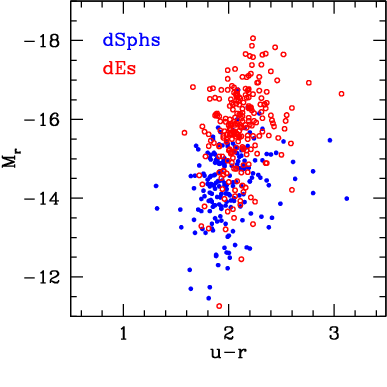}
\caption{Color-magnitude diagram  of early-type dwarf galaxies. Dwarf spheroidal galaxies are represented by filled blue circles and dwarf elliptical galaxies are plotted by open red circles.
}
\end{figure}
%\end{verbatim}%%%%%%%%%%%%%%%%%%%%%%%%%%%%%%%%%%%%%%%%%%%%%%%%%%%%%%%

\subsection{Data}

 We used the SDSS spectra of the selected dSphs and dEs which were downloaded from the SDSS DR7. The SDSS spectra are obtained by fibers of  $3^{\prime\prime}$ in diameter placed at the focal plane of the 2.5 m telescope of the Apache Point Observatory.  The SDSS spectrograph use 320 fibers with exposure time of 45 minutes or longer to achieve a fiducial signal-to-noise ratio. The spectra cover a wavelength range of 3800 to 9200 \AA  with a mean spectral resolution of $\lambda/\Delta \lambda \sim 1800$. The wavelength and flux are calibrated using pipeline developed by the SDSS team.  Since the average effective radius ($R_{e}$) of the sample galaxies is $\sim12^{\prime\prime}$ \citep{seo22},  the spectra we use in the present study reflect the stellar populations in the central regions of sample galaxies, $\sim13\%$ of $R_{e}$. We could obtain most of the spectra from SDSS DR7 but for a small fraction of sample galaxies ($\lesssim5\%$), we did not find relevant spectra.  We present the spectrum of one of dSphs whose  $r-$ magnitude is similar to the median values of the sample galaxies in Figure 3.
 The spectrum shows the typical features of early type dwarf galaxies. We use the observational data listed in the CVCG such as distance, luminosity ($M_{r}$), color ($u-r$),  morphological type as well as coordinates and redshifts.

%%%%%%%%%%%%%%%% Fig3  (Sp291664.png %%%%%%%%%%%%%
\begin{figure}
\centering
\includegraphics[width=0.52\textwidth]{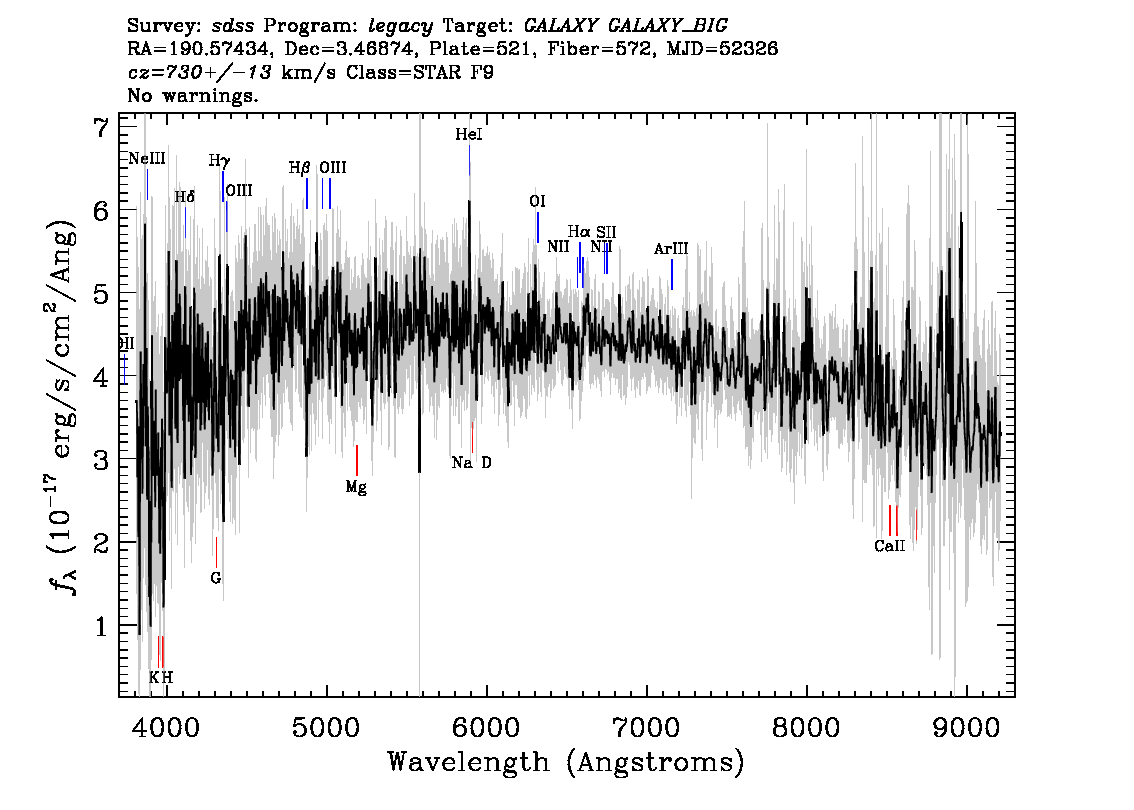}
\caption{Spectrum of VCC 1917 adopted from SDSS DR7. Some absorption features are designated.
}
\end{figure}
%\end{verbatim}%%%%%%%%%%%%%%%%%%%%%%%%%%%%%%%%%%%%%%%%%%%%%%%%%%%%%%%

\section{Star Formation Histories}
\subsection{Methods}

We applied STARLIGHT \citep{cidF05} to the SDSS spectra of 200 dSphs and 234 dEs to
obtain the most probable mix of stellar populations as a function of stellar age and metallicity. The STARLIGHT, a synthesis code, is well described by \citet{cidF04, cidF05} and a recent application is found in \citet{rif21}. It fits an observed spectrum with
the model spectrum derived from a combination of N simple stellar population (SSP) based on the evolutionary synthesis models of \citet[BC03]{bc03}. BC03 provides spectral evolution of SSPs of various metallicities (Z = 0.0001, 0.0004, 0.004, 0.008, 0.02, and 0,05) at ages between $1\times10^{5}$ and $2\times10^{10}$ yr with a resolution of 3 \AA  across the wavelength range from 3200 to 9500 \AA. They used the IMF of Chabrier (2003) with lower and upper mass cut-offs $m_{L} = 0.1 M_{\odot}$ and $m_{U} = 100 M_{\odot}$. The basic stellar evolutionary tracks are those from Padova groups complemented by Geneva groups with the STELIB library \citep{leb03}. 

In STARLIGHT, stellar motions projected to the line-of-sight are modeled by a Gaussian distribution ($G$) centered at radial velocity $v$ with velocity dispersion $\sigma$. The extinction due to foreground dust is taken into accounted using the V-band extinction $A_{V}$. Thus, the model spectrum
$M_{\lambda}$ is expressed as
\begin{equation}
  M_{\lambda}=M_{\lambda_{0}}(\sum_{j=1}^{N_{\ast}} x_{j} b_{j, \lambda} r_{\lambda}) \otimes G(v_{\ast},\sigma_{\ast})
\end{equation}

\noindent{where $b_{j, \lambda}$ is the $j$th SSP spectrum normalized at $\lambda_{0}$,
$r_{\lambda}=10^{-0.4(A_{\lambda}-A_{\lambda_{0}})}$, $M_{\lambda_{0}}$ is
the synthetic flux at the normalization wavelength $\lambda_{0}$, and $x_{j}$ is
the fractional contribution of the SSP for $j$th population that have
age $t_{j}$ and metallicity $Z_{j}$. The best fitting model is determined by
selecting a model that minimizes the $\Xi^{2}=\Sigma[(O_{\lambda}-M_{\lambda})w_{\lambda}]^{2}$ where $O_{\lambda}$ is observed spectrum and $w_{\lambda}$ is the inverse of error applied (see \citet{cidF04} for detailed description).

Since STARLIGHT compares the model spectrum at rest frame, we corrected for the
interstellar reddening and redshift before resampling the observed spectrum.
The reddening correction was made by multiplying $10^{0.4A_{\lambda}}$ to
the observed spectrum where $A_{\lambda}$ is interstellar extinction at
$\lambda$. We derived $A_{\lambda}$ by using the reddening law of \citet{ccm89}
with E(B-V) obtained from the dustmaps \citep{sfd98} by assuming the total-to-selective
extinction ratio ($R_{V}$) of 3.1.
After reddening correction, we shifted the observed wavelength to the rest
frame wavelength using the relation $\lambda_{rest} = \lambda_{obs}/(1+z)$
where $z$ is the redshift. We also applied the dimming of the flux due to
cosmic expansion by multiplying $(1+z)^{3}$ to the reddening corrected flux.
We resampled the corrected spectrum with a sampling width of
$\delta\lambda$ = 1 \AA  $ $  by applying a linear interpolation between the two nearest
data points for the STARLIGHT input spectrum.

For the analysis of the star formation histories, STARLIGHT provides
luminosity fraction ($x_{j}$) and mass fraction ($\mu_{j}$) for $j$-th
stellar population as a function of stellar age.
The $x_{j}$ and $\mu_{j}$ are divided into six stellar
metallicities (Z = 0.0001, 0.0004, 0.004, 0.008, 0.02, and 0.05), It assumes
[$\alpha$/Fe] = 0. The output of STARLIGHT gives two mass fractions,
$\mu_{ini}$ and $\mu_{cor}$,
which represent initial mass and mass corrected
for the mass returned to the interstellar medium, respectively. We used the
$\mu_{cor}$ for the mass fraction of stellar populations while we used the
$\mu_{ini}$ to derive the star formation rates.

\subsection{Mock spectra with noises from a Gaussian pertubation}

We applied STARLIGHT to the mock spectra to test the robustness of the
stellar populations we derived. We constructed the mock spectra using the
model fluxes from the STARLIGHT output which fits the observed spectra of
sample galaxies. The flux of a mock spectrum is calculated as 
\begin{equation}
F(\lambda) = F_{model}(\lambda) + F_{model}(\lambda)/(S/N) \times G(\lambda)
\end{equation}

\noindent{where $G(\lambda)$ is a number randomly drawn from a Gaussian normal 
distribution $N(0,1^2)$ at wavelength $\lambda$ (Eq. 6 of \citet{wan22}).
We used four signal-to-noise ratios ($S/N$) of 5, 10, 20, and 30, which well
cover the $S/N$ of the observed spectra of sample galaxies.
The total number of mock spectra is 1736 since four mock spectra were
constructed for each model flux of 434 early-type dwarf galaxies.}

%%%%%%%%%%%%%%%% Fig4  (Zage_mock.eps %%%%%%%%%%%%%
\begin{figure}
\centering
\includegraphics[width=0.45\textwidth]{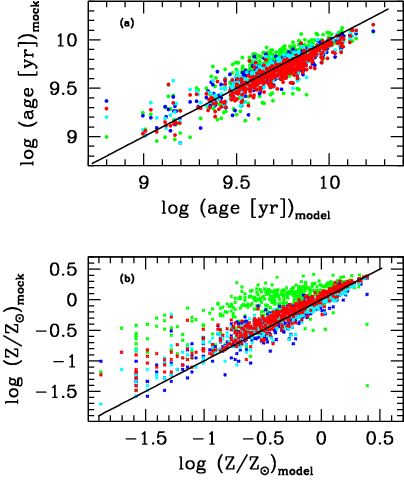}
\caption{Mock test with $S/N$ = 5, 10, 20, 30. The colors are assigned to 
distinguish $S/N$ : 5 (green), 10 (blue), 20 (cyan), and 30 (red).
The black solid lines are one-to-one relation lines.
	}
\end{figure}
%\end{verbatim}%%%%%%%%%%%%%%%%%%%%%%%%%%%%%%%%%%%%%%%%%%%%%%%%%%%%%%%

Figure 4 shows the mean ages and metallicities of mock galaxies compared with
those of model galaxies whose stellar populations were determined from the
observed spectra of sample galaxies using STARLIGHT. As shown in Figure 4-(a),
the method we employed to determine the stellar populations seems to well
reproduce the ages of mock galaxies of which spectra were perturbed by noises
proportional to ${S/N}^{-1}$. The metallicities of mock galaxies are also well
reproduced except for those from mock spectra with $S/N$ = 5. As shown in
Figure 4-(b), they are overestimated by $\sim0.3$ dex for solar metallicity
and $\sim0.7$ dex for metallicities of log$(Z/Z_{\odot}) < -0.5$. It implies
that the metallicity determination is more affected than the ages by spectral noises. It seems to be partially due to incomplete templates in BC03 SSPs
and the resemblance of the SSP spectra for the old metal-poor populations in BC03.

However, the ill reproduction of metalliciies from the mock spectra that
have S/N = 5 is not a severe problem for the present study because most SDSS
spectra of the sample galaixes have S/N greater than 5.
It is also worth to note that the
fluxes at wavelengths with large errors are masked during the fitting process
in STARLIGHT so that the best fitted model fluxes are not much affected by the
large errors in the observed fluxes which result in the small S/N.

\subsection{Stellar Mass}

The stellar masses of sample galaxies are calculated from the model fluxes determined by STARLIGHT using the distances in the CVCG. They are mostly in the range of  $10^{6}$ $\sim$ $10^{8}$ $M_{\odot}$. But they are not the total stellar masses of galaxies because they are calculated from the model fluxes fitted to the observed spectra through the $3^{\prime\prime}$ fiber which covers a small fraction of a galaxy image. In order to obtain the total stellar mass, we applied aperture correction ($AC$) calculated as
\begin{equation}
	AC= \frac{2\times \int_0^{R_{e}} f(r) dr} { \int_0^{R_{f}} f(r) dr}
\end{equation}
\noindent{where	$R_{e}$ and $R_{f}$ are the effective radius and the fiber radius, respectively, and {\it{f(r)}} is the S\'{e}rsic profile \citep{ser68}. We use the $R_{e}$ and  S\'{e}rsic index ($n$) that were determined by \citet{seo22}. For galaxies with unknown $R_{e}$ and $n$, we use the mean $R_{e}$ and $n$ derived for dSphs and dEs. They are $R_{e}\approx12^{\prime\prime}$ and $n \approx 0.9$ for dSphs and $R_{e}\approx11^{\prime\prime}$ and $n \approx 1.3$ for dEs, respectively. The number of galaxies with unknown S\'{e}rsic parameters is $\sim10\%$ of the sample galaxies. }The stellar masses of dSphs are on average $4$ times smaller than those of dEs, which corresponds to a magnitude difference of $\Delta m =1.5$ if we assume the same mass-to-luminosity ratio for both types. This magnitude difference is consistent with the magnitude difference shown in Figure 2. 

Figure 5 shows a comparison between the present estimates of the total stellar mass and those from SDSS DR12. The total stellar masses provided by SDSS DR12 were derived from multi-band photometric images following \citet{kau03} and \citet{bri04}. About $10\%$ of sample galaxies are missed in the stellar mass table of SDSS DR12. In particular, $\sim20\%$ of dSphs are missed in the SDSS DR12. There is a good correlation between the two sets of stellar masses. There are some outliers in the stellar masses from SDSS DR12 as well as the present estimates. We suppose that the outliers are mostly caused by large errors in the distances of galaxies because most of the distances are derived from the redshifts. In particular. the stellar masses larger than $10^{10}M_{\odot}$ are thought to be due to the larger distances applied to these galaxies. However, the tight correlation for the majority of sample galaxies is remarkable because the two approaches are completely independent. This confirms the robustness of the population synthesis applied in this study.

%%%%%%%%%%%%%%%% Fig5  (smass.eps %%%%%%%%%%%%%
\begin{figure}
\centering
\includegraphics[width=0.4\textwidth]{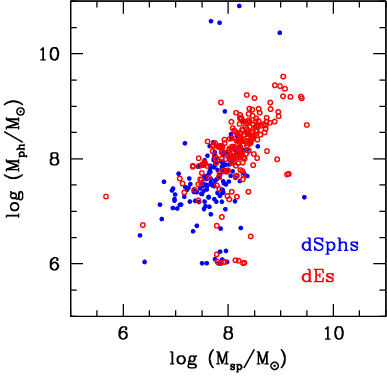}
\caption{Comparison of the stellar mass from STARLIGHT corrected for fiber aperture and the stellar mass from SDSS DR12. The horizontal axis for the stellar mass from STARLIGHT (M$_{sp}$) and the vertical axis for the stellar mass from SDSS DR12 (M$_{ph}$). Dwarf spheroidal galaxies are represented by filled blue circles and dwarf elliptical galaxies are plotted by open red circles. 
	}
\end{figure}
%\end{verbatim}%%%%%%%%%%%%%%%%%%%%%%%%%%%%%%%%%%%%%%%%%%%%%%%%%%%%%%%

\subsection{Luminosity and Mass Fraction}

Figure 6 and 7 show the mean luminosity and mass fractions of the stellar populations as a function of stellar age for dSphs and dEs, respectively. We applied sigma clipping that excludes galaxies deviate more than $3 \sigma$ from the mean. The number of rejected galaxies is less than $\sim3\%$ of the whole sample.  The SFHs of 200 dSphs and 234 dEs are provided in supplemented material as a 
tabular form, respectively. Some galaxies formed the majority of stars at early epochs of log ($t_{L}$) $\gtrsim10$ where $t_{L}$ is the lookback time in unit of yr, while others form most stars at later times, log ($t_{L}$) $ < 10$. The general trend of SFHs of the early-type dwarfs is the multiple bursts of star formation with decreasing strength toward the present time. The active periods of star formation ended just after log ($t_{L}$) $\approx 9$ and resumed at log ($t_{L}$) $\approx 8$ with much more reduced strength. The period of quenched star formation between log ($t_{L}$)  $\approx9$ and $8$ is frequently observed in the SFHs of the early-type dwarfs in the LG \citep{tol09}.
The recent star formation, which is virtually negligible in stellar mass fractions, is marginally noticeable sin luminosity fractions due to the high luminosity of young massive stars.  There is not much difference between dSphs and dEs for the SFHs after log ($t_{L}$) $\approx 8$.

The most notable feature of the SFHs of early-type dwarfs is the presence of two periods of active star formation. The first period is peaked at log ($t_{L}$) $=10$ and the second period occurs at log ($t_{L}$) $=9.4$. There are some differences in the characteristics of the double peaks between dSphs and dEs. In dSphs, the peak of the first period is $\sim6$ times higher than that of the second period in stellar mass fractions and $\sim3$ times higher in luminosity fractions, whereas similar strength in stellar mass fractions and reversed fractions in luminosity for dEs. It is apparent that majority of stars in dSphs are the stars formed in the first period. They contribute $\sim80\%$ of the present stellar mass and $\sim55\%$ of the present luminosity. More than $80\%$ of their mass is contributed by the stars older than $\sim10$ Gyr. The difference in the strength of the initial bursts is thought to be due to the difference in the effect of stellar feedback which depends on the mass of galaxies. In our sample of early-type dwarfs, dEs are, on average, 4 times more massive than dSphs. 

The difference in the fraction of stars in the second period of active star formation between dSphs and dEs is also related to the mass dependent stellar feedback. The stars formed in the second period of active star formation in dSphs is only half of that in dEs. This difference is thought to be caused by the different influence of stellar feedback due to their mass difference.  The strong suppression of star formation after the initial starbursts in dSphs is caused by supernova feedback which heats and expels the gas left to the halo.
The almost complete removal of gas in less massive dSphs is due to the explosive
starbursts in the early phase of galaxy formation. The lower mass of dSphs also provides a favorable condition for gas removal in the low mass dSphs.  The reionization feedback also plays a considerable role for the gas removal in low mass dSphs because reionization feedback is more effective in low mass galaxies \citep{daw18}.

The general behavior of star formation after the second period of star formation in dSphs is similar to that of dEs. It is characterized by a short period of quenched star formation until log ($t_{L}$) $\approx 8$ followed by increasing star formation activity toward the present time.  However, the total percentage of mass contribution by stars younger than $\sim0.1$ Gyr is less than $5\%$.
The almost quenched star formation between log ($t_{L}$) $\approx 9$ and $0$ resulted in the absence of intermediate-age stars older than $\sim0.1$ Gyr.
This feature of a highly suppressed star formation of intermediate age stars is also observed
in some early-type dwarfs in the LG \citep{wei14a}. The suppression of
star formation is thought to be caused by stellar feedback such as supernova explosion and stellar wind. The cold gas of galaxies was
heated and easily expelled to the halo of early-type dwarfs.
After that, star formation was resumed and continued to the present time with
multiple starbursts to produce young stellar populations which contribute
$\sim5\%$ of the present luminosity and $\lesssim0.2\%$ of the stellar mass
in total. The resumed star formation at log ($t_{L}$) $\approx 8$ is thought to be
driven by the accretion of cold IGM because the metallicities of stars
formed just after the period of quenched star formation can be explained if
we assume accretion of metal free IGM as described below.

%%%%%%%%%%%%%%% Fig6 (dSphLxMc.eps) %%%%%%%%%%%%%%%%%%%%%%%%%%%%
\begin{figure}
\centering
\includegraphics[width=0.4\textwidth]{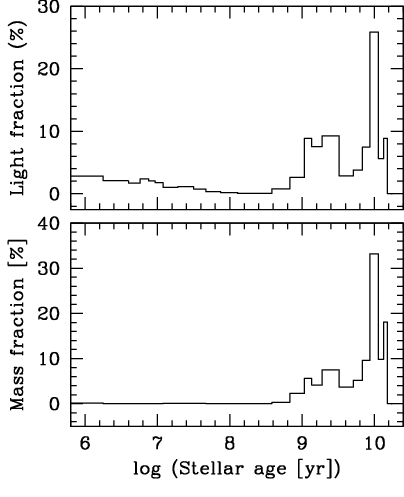}
\caption{Mean luminosity and mass fractions as a function of stellar age
for 200 dwarf spheroidal galaxies. }
\end{figure}
%\end{verbatim}%%%%%%%%%%%%%%%%%%%%%%%%%%%%%%%%%%%%%%%%%%%%%%%%%%%%%%

%%%%%%%%%%%%%%% Fig7 (dELxMc.eps ) %%%%%%%%%%%%%%%%%%%%%%%%%%%%
\begin{figure}
\centering
\includegraphics[width=0.4\textwidth]{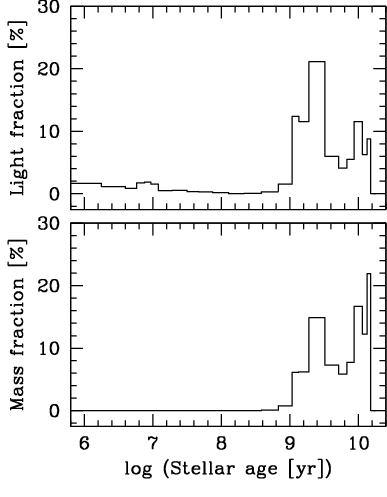}
\caption{Mean luminosity and mass fractions as a function of stellar age
for 234 dwarf elliptical galaxies. }
\end{figure}
%\end{verbatim}%%%%%%%%%%%%%%%%%%%%%%%%%%%%%%%%%%%%%%%%%%%%%%%%%%%%%%

\subsection{Metallicity Dependent Mass Fractions}

\subsubsection{Dwarf Spheroidal galaxies}

Figure 8 shows the mean mass fractions of the stellar populations of 200 dSphs, divided into 6 metallicity groups, as a function of stellar age. As described above for the derivation of mean luminosity and mass fractions, we applied a sigma clipping method to exclude outliers. The stellar metallicities we used are Z = 0.0001, Z = 0.0004, Z = 0.004, Z = 0.008, Z = 0.02, and Z = 0.05. The most pronounced feature of the SFH of dSphs is the explosive star formation at log ($t_{L}$) = 10 which produced $\sim40\%$ of the present stellar mass. More than $99\%$ of them are metal poor (Z = 0.0004) stars and the rest have metallicities of Z = 0.0001 and Z = 0.004  with negligible fractions. 

The first period of star formation continued to form
metal-poor stars until log ($t_{L}$) $\approx 9.8$ and contributes $80\%$ of the present stellar mass.
In terms of luminosity, the metal poor stars formed in the first period of star formation contribute  $54\%$ to the present luminosity. Since the majority of the oldest stars formed at log ($t_{L}$) $\gtrsim 10$ have a metallicity of Z = 0.0004, which corresponds to [Fe/H] = -2.6 assuming the solar abundance of 0.01524 \citep{bre12},
the first generation stars in dSps are assumed to be formed from pre-enriched gas.
The metals from stellar evolution enriched the gas left, but a larger fraction of metals was thought to be ejected into the halo \citep{eme19}. We expect some halo metal fell back to the
nuclear region to enrich the gas left further.

The metallicity distribution of the stars formed in the second period of the active star formation is much different from that of the first period. A variety of metallicities from Z = 0.0001 to Z = 0.05 are involved in the stars formed in the second period.  It begins with Z = 0.004 at the beginning of the second period and increases with decreasing stellar age nearly monotonically. At the end of the second period at log ($t_{L}$) $\approx 8.9$, the metallicity of stars becomes Z = 0.05. As the stellar ages become younger, the metallicity of stars increases to become Z = 0.05 at the end of the second period. The dominant stellar populations of the second period of active star formation are stars of intermediate metallicity (Z = 0.004 and Z = 0.008), with a slightly higher fraction of Z = 0.008.

The stellar populations formed in the second
period of star formation contribute $20\%$ of the present stellar mass. Among the stellar
populations formed in the second period, the mass of the extremely metal-rich stars is $\sim4\%$ of the
present stellar mass. The presence of a considerable amount of metal-rich populations
suggests that a rapid chemical evolution had occurred in the majority of
dSphs.

The third period of star formation, from log ($t_{L}$) $\approx 8$ to the present time, shows several episodes of a starburst with very weak intensity. The stars in this period have a variety of metallicity, from Z = 0.0001 to Z = 0.05 with a large contribution of extremely metal poor stars formed at log ($t_{L}$) $\approx 6.7$. These bursts produced stars that contribute negligible fractions ($\lesssim 0.1\%$) to
the present stellar mass.

Since there is negligibly small amount of extremely metal-poor stars formed in the
first and second periods of star formation in dSphs, the extremely metal-poor stars formed recently could not be formed from the gas that fell into the potential well of the dark matter halo in the early
phase of its formation. Rather, they were formed from the gas that came in
recently from the surrounding cold IGM because the metallicity of the gas in
the nuclear regions of a galaxy increases monotonically unless pristine
IGM is accreted onto the nuclear regions. The presence of HI gas around
galaxies is well established \citep{dan08} with more frequent in field galaxies than in group
galaxies \citep{wak09}.
However, there are two possibilities for the origin of the gas out of which
the stars that have metallicities higher than Z = 0.0001 form.
One is that it is the gas ejected
into the halo of dSphs which cooled and settled down into
the nuclear regions of the galaxy to form stars. The variety of metallicity
for the young stellar populations can be explained if the metals ejected into
the halo do not distribute uniformly. The other is that it is the cold IGM
which is inhomogeneously mixed with the metals in the halo of the galaxy.
Since dSphs are mostly consisted of old stellar populations \citep{gal94, gre03} and they do not have
cold gas out of which stars form. These young stellar populations are thought to be formed from
the accreted IGM.  In any case, the accretion of IGM played an important role because it seems to be metal free, which is required to produce the extremely metal-poor stars.

%%%%%%%%%%%%%%% Fig8 (csfh_Mc_stackbar_dSphRev.png) %%%%%%%%%%%
\begin{figure}
\centering
\includegraphics[width=0.42\textwidth]{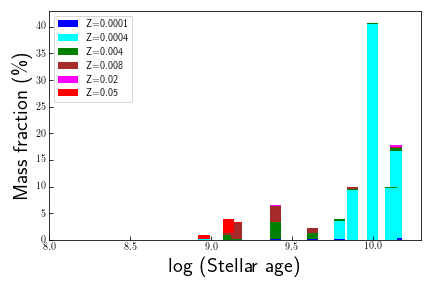}

\caption{Mean mass fractions of stellar populations as a function of stellar age
for 200 dwarf spheroidal galaxies. Stellar populations are divided by their
metallicities.}
\end{figure}
%\end{verbatim}%%%%%%%%%%%%%%%%%%%%%%%%%%%%%%%%%%%%%%%%%%%%%%%%%%%%%%

\subsubsection{Dwarf Elliptical galaxies}

The mean SFH of dEs, divided according to stellar metallicity as a function of stellar
age, is presented in Figure 9. The characteristics of SFH, distinguished by stellar metallicities, are much different from that of dSphs shown in Figure 8. First of all, the early bursts occurred at log ($t_{L}$) $\gtrsim 10.1$ is very explosive that made rapid metal enrichment of the interstellar medium from which generations of stars form. The explosive star formation of the initial burst is already seen in Figure 7, but here we see a rapid chemical evolution at the earliest phase of galaxy formation. The explosive bursts at log ($t_{L}$) $\gtrsim 10.1$ produced four kinds of metallicity, from Z = 0.0004 to Z = 0.05 but the contribution by the extremely metal rich (Z = 0.05) stars is negligibly small, $\sim0.3\%$. It seems difficult to consider the presence of extremely metal rich stars as a real feature since it requires an unusually large number of cycles of star formation. If it is a real feature, there is an extremely inhomogeneous mixing of metals to make extremely metal rich stars. Pre-enrichment may have some role for rapid metal enrichment. The fraction of metal rich (Z = 0.02) stars is considerable enough to confirm a rapid chemical evolution in dEs although the majority of stars formed in the first period of active star formation are metal poor stars.

The chemical evolution of dEs during the second period of star formation is similar to that of dSphs, that is, it started with intermediate metallicities (Z = 0.004 and Z = 0.008) and ended with extremely metal rich stars. The most dominant metallicity in the second period of star formation is higher intermediate metallicity (Z = 0.008) which amounts to more than half of stars formed in the second period. The larger contribution of stars with Z = 0.008 than dSphs is due to the more rapid enrichment in dEs. Owing to rapid enrichment, extremely metal rich stars are found over a large range of stellar ages. Virtually, all generations of stars formed in the first and second periods of star formation have extremely metal rich stars, although they contribute to the present stellar mass negligibly small except for the stars formed in the late phase of the second period at log ($t_{L}$) $\approx 9.1$.

%%%%%%%%%%%%%%% Fig9 (csfh_Mc_stackbar_dERev.png)  %%%%%%%%%%
\begin{figure}
\centering
\includegraphics[width=0.45\textwidth]{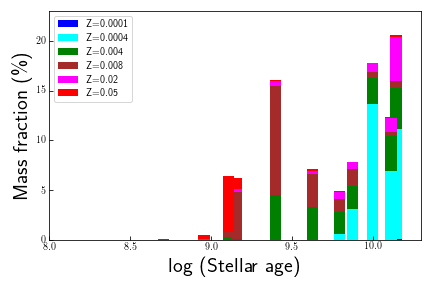}

\caption{Mean mass fractions of stellar populations as a function of stellar age
for 234 dwarf elliptical galaxies. Stellar populations are divided by their
metallicities.}
\end{figure}
%\end{verbatim}%%%%%%%%%%%%%%%%%%%%%%%%%%%%%%%%%%%%%%%%%%%%%%%%%%%%%%

Aside from the pronounced differences between dEs and dSphs, one notable difference in SFHs between dEs and dSphs is observed in the fractions of the extremely metal poor (z = 0.0001) stars.
The contribution of the extremely metal poor star to the present stellar mass is less than $1\%$ in dEs while that in dSphs is slightly larger than $2\%$. This difference is more pronounced in the fractions of the oldest populations which contribute the highest fractions to the extremely metal poor stars. The extremely metal poor stars in the oldest population of dEs contribute only $0.2\%$ of the present stellar mass while those of dSphs contribute $0.8\%$.

The metallicity distribution of stars formed after the second period of
star formation is not much different from that of dSphs.
There is a very weak star formation after the end of the second period at log ($t_{L}$) $\approx 8$. Only  $\lesssim0.2\%$ of the present mass was formed after the second period.

\section{Cumulative Star Formation Histories}
We derived the cumulative star formation history (cSFH) of dSphs and dEs to examine the mass assembly history of early-type dwarf galaxies. We analyzed two parameters related to the cSFH. One is the epoch of maximum starburst which shows the formation time of the dominant stellar population. The other is the quenching time which is defined as the lookback time when most of the galaxy mass is assembled. There are some advantages to use cSFH rather than SFH because cSFH circumvents the uncertainties derived in the absolute SFHs \citep{jos21} as well as suppresses the noise caused by the limited resolution of stellar ages.

\subsection{Epoch of Maximum Star Formation}

It is well known that there is a bi-modality in the age of stellar populations of galaxies demonstrated by two peaks in the galaxy age distribution \citep{gal05, gal08}. One peak is caused by the old early-type galaxies and the other peak by young star forming galaxies. The distribution of stellar ages presented in Figures 6 - 9 strongly suggests a bi-modality of star formation in early-type dwarf galaxies, especially in dEs. The reason for a weak bi-modality in the SFHs of dSphs seems to be due to feedback of supernova explosion which heats the cold gas to blow out  \citep{dek86}. Figure 10 shows the distribution of epochs of starbursts. It is clear that star formation in early-type dwarfs is not a single event but multiple episodes of starbursts. The first epoch of starbursts occurs at log ($t_{L}$) $\gtrsim 10.1$ for both dSphs and dEs. This epoch includes the first and second bursts shown in Figures 6 - 9.  The fraction of galaxies which form stars at the first epoch of starburst is about $30\%$ of dSphs and $43\%$ of dEs, respectively. The first epoch of starburst is the second-high peak in dSphs while it is the highest peak in dEs. The $86\%$ of dSphs and $76\%$ of dEs result in starbursts during the first period of star formation. The highest peak of the second period of star formation in dSphs occurs at log ($t_{L}$) $\approx 9.4$. This peak is made of 10 dSphs only. Thus, the bi-modality of star formation in dSphs is very weak compared to that of dEs of which 44 galaxies ($\sim20\%$) have starbursts at this epoch. There are some bursts of star formation after the second period of star formation in dSphs, but it occurs in $\sim2\%$ of dSphs. It is greatly contrasted with the bi-modality of star formation commonly observed in starburst galaxies \citep{cidF18}.

%%%%%%%%%%%%%%% Fig10 (peakepoch.eps) %%%%%%%%%%%%%
\begin{figure}
\centering
\includegraphics[width=0.4\textwidth]{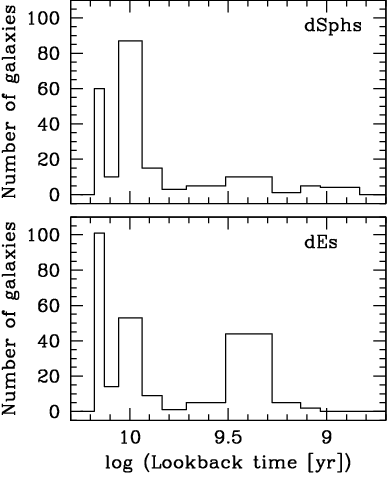}
\caption{Frequency distribution of the epochs of maximum starbursts. }
\end{figure}
%\end{verbatim}%%%%%%%%%%%%%%%%%%%%%%%%%%%%%%%%%%%%%%%%%%%%%%%%%%%%%%

%%%%%%%%%%%%%%% Fig11 (qtauhist.eps) %%%%%%%%%%%%%%%%%
\begin{figure}
	\centering
	\includegraphics[width=0.4\textwidth]{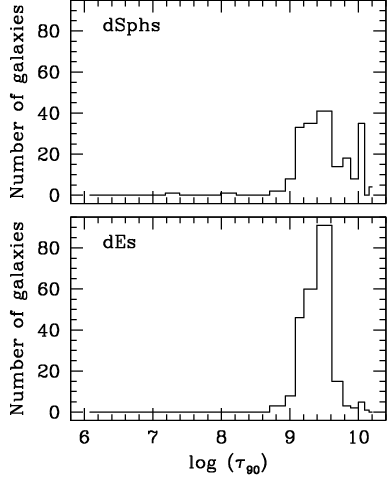}
	\caption{Frequency distribution of quenching time $\tau_{90}$. }
	
\end{figure}
%\end{verbatim}%%%%%%%%%%%%%%%%%%%%%%%%%%%%%%%%%%%%%%%%%%%%%%%%%%%%%%

\subsection{Quenching Epoch}
In past studies of the SFHs of galaxies, it is a common practice to present the characteristics of SFHs in a graphical form such as the figures in the previous sections. \citet{wei14a, wei14b, wei15} introduced a parameter, quenching time $\tau_{90}$, which is defined as the lookback time at which 90\% of stellar mass is formed, to quantify the cSFHs of galaxies. Figure 11 shows the histograms of $\tau_{90}$ for the 200 dSps and 234 dEs.  As expected from the difference in SFHs of dSphs and dEs shown in the above, there is a clear difference in the distribution of  $\tau_{90}$. In particular, a significant fraction of dSphs have log ($\tau_{90}$) > $10$ while only a negligible fraction of dEs have log ($\tau_{90}$) > $10$. The extremely early quenching of $\sim20\%$ of dSphs is thought to be due to small mass of these galaxies since low mass dwarfs are known to be quenched earlier \citep{dig19, gar19, jos21}. Besides it, there are  two other differences between the dSphs and dEs. One is the highly concentrated distribution of $\tau_{90}$ in dEs compared with the broad distribution of $\tau_{90}$ in dSphs. In both dSphs and dEs, the peak, i.e.,   the maximum frequency, occurs at log ($t_{L}$) $\approx 9.7$, but the peak height is $\sim2$ times higher in dEs than dSphs. The other is that virtually no dEs have log ($\tau_{90}$) less than 8.  It seems highly probable that the dSphs with log ($\tau_{90}$) > 10 are bona fide primordial objects. The paucity of such early quenched dEs may imply that most of dEs are not primordial objects.

\subsection{Dependence on Morphology, Stellar Mass and Environment}

We investigated the dependence of cSFHs on the morphology, stellar mass, and environment of early-type dwarfs. We used the present estimates of the total stellar mass and the local background density derived from the data in the  CVCG.  

\subsubsection{Morphology}
Figure 12 shows the cSFHs of early-type dwarfs as a function of lookback time, grouped by the two morphological types, dSph and dE. We selected galaxies with the local background density around the median density ($\Sigma_{m}$), i.e., $\Sigma_{m} - 1\sigma < \Sigma < \Sigma_{m} + 1\sigma$ to avoid the effect of the local background density on the cSFHs. We present the mean cSFHs and $\pm 1\sigma$ boundaries. There is a significant difference in the cSFHs between dSphs and dEs from log ($t_{L}$) $\approx 10$ to log ($t_{L}$) $\approx 9.5$. It amounts to $\sim 0.8\sigma$ from the mean cSFH of dEs at log ($t_{L}$) $\approx 9.8$.  This difference is mainly due to the explosive bursts of star formation in dSphs at log ($t_{L}$) $\approx 10$, as shown in Figure 8. It causes a rapid increase in cSFH of dSphs after log ($t_{L}$) = 10. Owing to the rapid assembly of stellar mass in early-type dwarfs, it becomes 62\% of the present stellar mass in dSphs and 52\% of the present stellar mass in dEs at log ($t_{L}$) = 10. 

The evolution of cSFHs of dSphs and dEs shows three characteristic features. The first feature is the very steep slopes of dSphs and dEs at log ($t_{L}$) $\gtrsim 10$. Both types have nearly the same slopes in this period. The second feature is the significant difference in the cSFHs in the second period of star formation from log ($t_{L}$) $\approx 9.7$ to log ($t_{L}$) $\approx 9$. There is a break in the cSFH of dSphs while dEs have a nearly constant slope. The slope before the brake at log ($t_{L}$) $\approx 9.8$ is similar to that of dEs while the slope after the brake is shallower than that of dEs. The cSFHs are nearly constant after log ($t_{L}$) $\approx 9$ for both types due to quenching at log ($t_{L}$) $\approx 9$.

%%%%%%%%%%%%%%% Fig12 (csfh_dEdSph_med_1Rev3.png) %%%%%%%%%%
\begin{figure}
\centering
\includegraphics[width=0.4\textwidth]{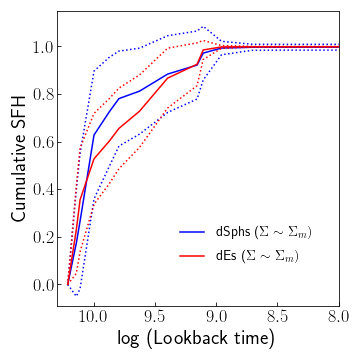}
\caption{Cumulative SFHs of dSphs and dEs. We selected galaxies with the local background density, $\Sigma_{m} - 1\sigma < \Sigma < \Sigma_{m} + 1\sigma$ where $\Sigma_{m}$ and $\sigma$ is the median and standard deviation of the local background density, respectively. We plot the mean cSFHs of dSphs (blue solid line) and dEs (red solid line) together with 1$\sigma$ boundaries (dotted lines).}

\end{figure}
%\end{verbatim}%%%%%%%%%%%%%%%%%%%%%%%%%%%%%%%%%%%%%%%%%%%%%%%%%%%%%%

As can be inferred from the $1\sigma$ upper boundary of the cSFH of dSphs in Figure 12, a significant fraction of dSphs make $\sim90\%$ of their stars during the first period of star formation. This is one of the reasons why we think most of dSphs are primordial objects while a significant fraction of dEs are transformed ones.

%%%%%%%%%%%%%%% Fig13 (csfh_dEdSph_morph_mass.png) %%%%%%%%%%%
\begin{figure}
	\centering
	\includegraphics[width=0.4\textwidth]{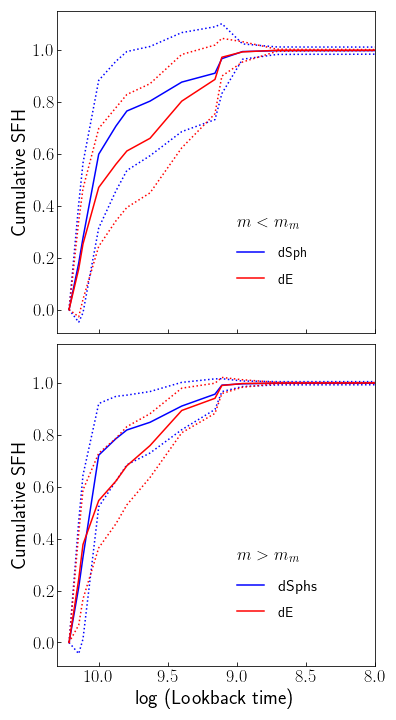}
	\caption{
	Cumulative SFHs of dSphs and dEs for the two stellar mass groups 
	divided by the median mass of the whole sample which 
	is $5\times10^{7}M_{\odot}$. The cSFHs of galaxies in the low mass group are plotted in the upper panel and those of the high mass group are presented in the lower panel. The cSFHs of different morphology are distinguished by colors: dSphs (blue), dEs (red).}
\end{figure}
%\end{verbatim}%%%%%%%%%%%%%%%%%%%%%%%%%%%%%%%%%%%%%%%%%%%%%%%%%%%%%%

%%%%%%%%%%%%%%% Fig14 (csfh_dEdSph_morph_mass.png) %%%%%%%%%%%
\begin{figure}
	\centering
	\includegraphics[width=0.4\textwidth]{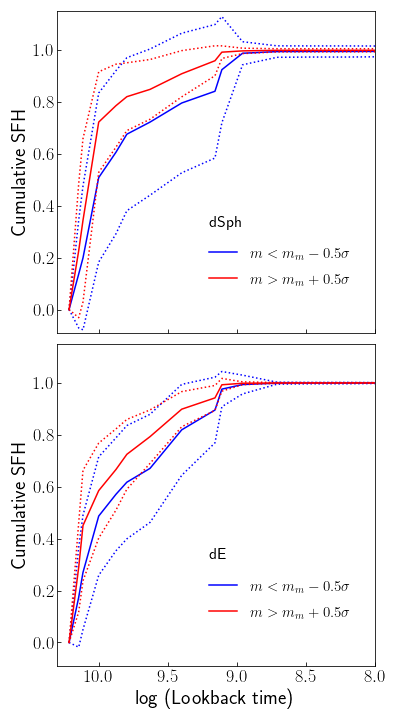}
	\caption{Cumulative SFHs of dSphs and dEs for the two stellar mass groups.
		Galaxies in the low mass group have total stellar mass ($m$) less than the median mass ($m_{m}$) - 0.5$\sigma$ and those in the high mass group have $m$ greater $m_{m}$ +  0.5$\sigma$. Here, $m_{m}$ is the median mass calculated for each type. Upper panel presents the cSFHs of dSphs
		and lower panel displays those of dEs.
	}
\end{figure}
%\end{verbatim}%%%%%%%%%%%%%%%%%%%%%%%%%%%%%%%%%%%%%%%%%%%%%%%%%%%%%%

\subsubsection{Stellar Mass}

As can be seen in Figure 5, it is apparent that dSphs and dEs have different stellar mass distributions. The distinction of cSFH of dSphs from that of dEs shown in Figure 12 may be caused by the different stellar masses of the two types. Thus, it is of interest to see which is the main cause of the different cSFHs of the two types. There are two ways to separate the effects of morphology and stellar mass. One is to examine the dependence of cSFHs of the two types by constraining the stellar mass, and the other is to see dependence of cSFHs on the stellar mass by fixing the morphological type. To do this, we used the stellar masses of galaxies determined by population synthesis using STARLIGHT, corrected for the fiber aperture. 

Figure 13 shows the dependence of cSFHs of early-type dwarfs on their detailed morphology. We divide the sample galaxies in two mass groups using the median stellar mass of the whole sample ($\sim 5\times10^{7}M_{\odot}$). It is clear that the cSFH of dSphs is much different from that of dEs, regardless of their masses. Of the two mass groups, the high mass group shows more significant difference than the low mass group as the cSFHs of the two types differ about $1\sigma$ from others, which results in the probability of the K-S test as small as $p = 0.03$. The feature seen in Figure 12, i.e., more rapid star formation in dSphs than dEs, is confirmed here. The stellar mass dependence of cSFHs can be seen if we compare the cSFHs for the low mass group with those for the high mass group. The difference between low mass group and high mass group is more pronounced in the cSFH of dSphs than dEs. It suggests that the SFHs of the low mass early-type dwarfs are more affected by the stellar mass than that of the high mass early-type dwarfs. Since morphology is closely related to the SFHs of early-type dwarfs regardless of their masses, together with a moderate dependence of cSFHs of dEs on stellar mass, morphology seems to be more important in the SFHs of early-type dwarfs than the stellar mass.  

Figure 14 shows the dependence of cSFHS of dSphs and dEs on the stellar mass. It is apparent that the cSFHs of both types depend on the stellar mass with somewhat stronger dependence in dSphs. Here we used two mass groups separated by $1\sigma$ difference in their total stellar mass. The cSFH of dSphs in the high mass group is $\sim60\%$ higher than that of the low mass group until log ($t_{L}$) = 10. The difference in the cSFHs between the low mass dSphs and the high mass dSphs becomes largest at log ($t_{L}$) $\approx 10.1$, and decreases smoothly until log ($t_{L}$) $\approx 8.7$. Thereafter, the two cSFHs are almost the same. The cSFH of the low mass dSphs is similar to the lower $1 \sigma$ boundary of the cSFH of the high mass dSphs. The K-S test shows that the two groups are significantly different with the probability of $p = 0.03$. Thus,  dSphs with significantly different stellar mass are likely to have different star formation histories. A similar but less pronounced difference is observed in the cSFHS of dEs. This means that, regardless of morphology, mass is thought to play an important role in the SFHs of early-type dwarfs. The main cause of the difference in the cSFHs between the two mass groups of early-type dwarfs is largely due to the initial explosive bursts of star formation in more massive galaxies. One of the reasons for the less active star formation in the low mass early-type dwarfs seems to be the supernova feedback that removes a large amount of gas left after the initial bursts of star formation. If we consider Figure 13 and Figure 14 together, morphology is, at least, as important as stellar mass for the SFHs of the early-type dwarfs.

\subsubsection{Local Background Density}

The environment of dwarf galaxies in connection with galaxy morphology has been explored by \citet{vdb94}, \citet{kar14}, and \citet{ann17}, respectively for the LG, the local volume within 10Mpc, and the local universe $z\lesssim0.01$. 
We calculated the local background density ($\Sigma$) by the $n$th nearest neighbor method with $n = 5$ and normalized by their mean values. We used the linking distance of $LD = 1$ Mpc and linking velocity of $\Delta V^{\ast}$ = 1000 km s$^{-1}$ adopted by \citet{ann17}. 
Figure 15 shows the cSFHs as a function of the lookback time for dSphs and dEs in two density groups. As is the case for the dependence on the stellar mass, we used the median ($\Sigma_{m}$) and $\sigma$ to assign groups. As can be seen, the cSFH of dSphs and dEs in the high density regions is much different from that for the low density regions, for log ($t_{L}) > 9.2$.  It seems plausible that early-type dwarfs in the high density regions are assembled earlier than those in the low density regions because dynamical time scale such as gravitational collapsing time is inversely proportional to the square root of gas density, i.e., $\tau_{dyn} \sim \rho^{-1/2}$. Galaxies formed in the high density regions collapse more rapidly than those in the low density regions. The local density dependence observed in the cSFHs of dSphs and dEs is understandable if they are formed from the primordial gas clouds which collapse by gravitation. In the earliest phase of galaxy formation at log ($t_{L}$) $\gtrsim 10.1$, the gas cloud out of which the first generation of stars formed collapse more rapidly than those in the low density regions, regardless of their morphology. The initial explosive bursts of star formation observed in dEs seems to be the outcome of the fast collapse at log ($t_{L}$) $\gtrsim 10.1$

%%%%%%%%%%%%%%% Fig15 (csfh_dEdSph_mean_bden_1sigma.png) %%%%%%%%%%%%
\begin{figure}
\centering
\includegraphics[width=0.4\textwidth]{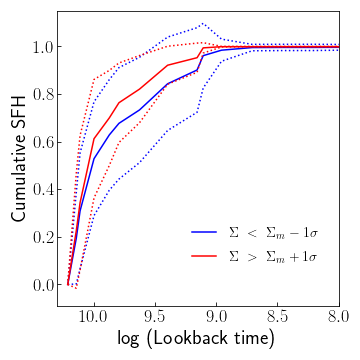}
	\caption{Cumulative SFHs of dSphs and dEs for the low and high background density ($\Sigma$). We selected galaxies with $\Sigma $ less than $\Sigma_{m}$ - 1$\sigma$ as the low density group and those with $\Sigma $ greater than $\Sigma_{m}$ + 1$\sigma$ as the high density group where $\Sigma_{m}$ and $\sigma$ are the median and standard deviation of the local background density of sample galaxies. The solid lines represent the mean cSFHs and the dotted lines are $\pm$ 1$\sigma$ boundaries, blue color for the low density group and red color for the high density group, respectively. }

\end{figure}
%\end{verbatim}%%%%%%%%%%%%%%%%%%%%%%%%%%%%%%%%%%%%%%%%%%%%%%%%%%%%%%

For dEs of primordial origin, the gas out of which stars formed in this period is the gas expelled to the halo by supernova feedback caused by the initial explosive bursts of star formation and accreted to the galaxy after being cooled. In this case, the low background density is more favorable to the gas cooling than the high density regions because interactions with the ambient hot intergalactic medium is less effective in the low density regions. If most of dEs are transformed from late-type galaxies, the opposite trend is expected because gas removal is more effective in the high density regions by ram pressure stripping \citep{gg72} which is thought to be the major mechanism to remove the gas from the late-type galaxies. Figure 16 shows the very feature that supports the transformation hypothesis as the origin of dEs although the background density dependence is not statistically significant. Since the difference due to different local background density at fixed morphology is much smaller than the differences due to stellar mass or morphology, it is difficult to exclude the possibility of the primordial origin of dEs.

%%%%%%%%%%%%%%% Fig16 (csfh_dEdSph_mean_bd_median.png) %%%%%%%%%%%%
\begin{figure}
\centering
\includegraphics[width=0.4\textwidth]{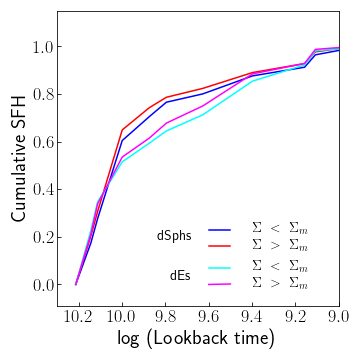}
\caption{Cumulative SFHs of dSphs and dEs, divided by the local background density ($\Sigma$). The low density group has $\Sigma < \Sigma_{m}$ and the high density group has $\Sigma > \Sigma_{m}$ where $\Sigma_{m}$ is the median $\Sigma$ of early-type dwarfs. The dSphs of the low density and high density groups are plotted by blue and red lines, respectively, whereas dEs are plotted by cyan and magenta lines, respectively for the low and high density groups. }
	
\end{figure}
%\end{verbatim}%%%%%%%%%%%%%%%%%%%%%%%%%%%%%%%%%%%%%%%%%%%%%%%%%%%%%%

\section{Discussion}

\subsection{Pre-enrichment and Supernova Feedback}
The distinction of dSphs from dEs may help to understand the origin of early-type dwarfs more clearly because their SFHs are significantly different, which implies different origins of dSphs and dEs. The different characteristics of the SFHs of the two sub-types of the early-type dwarfs seems to be closely related to the critical issues on the formation and evolution of galaxies, pre-enrichment and quenching time.

Pre-enrichment of interstellar medium (ISM) and IGM by supernovae from Pop III stars have been thought to be present by simulations \citep[e.g.,][]{bro09} and observations which show absence of extremely low metallicities below Z $ \approx10^{-3}$ Z$_{\odot}$ \citep{cow98, hel06}. The observations that showed a metallicity floor of
$\sim10^{-3}$ Z$_{\odot}$ include the metallicity of the Milky Way halo stars, the LG dSph galaxies \citep[e.g.,][]{hel06}, damped Ly${\alpha}$ absorption systems \citep{wol98}, and intervening IGM in the line of sight toward
quasars \citep[e.g.,][]{cow98}. The absence of extremely metal-poor (Z = 0.0001) stars among the oldest stellar populations in the majority of the early-type dwarfs (dSphs and dEs) reported here suggests pre-enrichment of the gas out of which the oldest stars formed. The pre-enrichment resulted in the metallicities of the oldest stellar
populations as Z = 0.0004. This level of pre-enrichment is in good agreement with recent simulations \citep{tas12, wis12} in which metals from Pop III supernovae enrich the ISM of the hosting halo and the surrounding IGM.

It is unclear whether the pre-enrichment was made before the collapse of gas into dark matter halo
or after the assembly of galaxies. However, the relatively higher metallicity of Z = 0.0004 for the first generation of stars in the early-type dwarfs favors in-situ pre-enrichment during the reionization.
This scenario is plausible because the inhomogeneous gas surface density could provide a favorable condition for the formation of Pop III stars in the high density pristine gas pockets \citep{tas12,pal14}.
However, pre-enrichment before galaxy assembly is not completely excluded
because the gas out of which a galaxy formed could be pre-enriched by Pop III stars in the neighboring minihalos. It is also possible that pre-enrichment happened in the subhalos that were merged into the host halo. The ages of the oldest stellar population of dSphs and dEs ($\gtrsim10$ Gyr) suggest that star formation started just after the end of reionization which is supposed to occur at z$\approx6$ \citep{fan06}.

The violent bursts in $\sim30\%$ of dSphs and $\sim35\%$ of dEs at log ($t_{L}$) $> 10$  imply that their mass is larger than the characteristic mass below which star formation is suppressed by the reionization feedback. For the dark matter halo, this characteristic mass $M_{halo}$ is $\sim10^{10} M_{\odot}$ \citep{fit17} which corresponds to the virial temperature of $\sim10^{4}$ K \citep{ben15}.  In terms of stellar mass ($M_{\ast}$), it is $\sim10^{6} M_{\odot}$ based on the extrapolated $M_{\ast}$ - $M_{halo}$ relation and
simulations \citep[references there in]{fit17}. These mass scales are
consistent with the upper limiting mass of ultra faint dwarfs ($\sim10^{5} M_{\odot}$) whose star formation was quenched by reionization feedback \citep{rod19}.

There is a signature of severe suppression of star formation in early-type dwarfs due to the feedback of stellar evolution such as supernova explosions in the initial bursts of star formation. This feedback seems to be strong enough to heat the gas left after the initial bursts of star formation and suppresses star formation somewhat before the second peak of star formation at log ($t_{L}$) $= 9.4$. Suppression of star formation after the first period of star formation also seems to be caused by the stellar feedback that heats gas and expels some of them. Removal of gas by supernova feedback seems to be more effective in dSphs because the fraction of stars formed in the second period of star formation is much smaller than those of dEs which show comparable fractions in the first and second period of star formation.

The present sample of early-type dwarfs shows gappy SFHs which are characterized by distinct quiescent periods of star formation \citep{wri19} more frequent than the LG dwarfs \citep{wei11}. In particular, dEs in the present sample show mostly gappy SFHs. Since dEs are, on average, more massive than dSphs, gappy SFHs are likely to be made for galaxies where the supernova feedback is strong enough to heats and expels the gas left to the halos of galaxies but not sufficient to remove the gas to the intergalactic matter. The gas ejected into the halos of galaxies accretes to the galaxies after cooling down and begins to collapse to form next generation of stars. The mass dependent gappy SFH is understandable because the feedback of supernova explosion is more effective in less massive galaxy.  In less massive dwarfs, the stellar feedback removes the gas left completely after the first period of star formation.

\subsection{Morphology as a fingerprint of the Origin of Early-type dwarfs}
From the analysis of SFHs of dSphs and dEs, it seems apparent that the morphology of early-type dwarfs is closely related to their SFHs. As shown in Figure 13, the cSFH of dSphs is clearly different from that of dEs, regardless of their mass. It implies that the morphology dependence of cSFHs seems to be most important in the SFHs of early-type dwarfs (Figures 13 and 14). Thus, dSphs are not merely a faint sub-sample of dEs but are distinct objects. The morphology difference seems to be due to their different origins. The majority of dSphs are likely to be primordial objects while most dEs are likely to be transformed ones. It is more likely that most faint dSphs are originated from the primordial objects some of which are collapsed before reionization while the majority of bright dEs are transformed from the late-type galaxies as implied by the embedded spiral arms in a number of dEs \citep{lis06, seo22}. However, it is highly plausible that some fractions of dEs are primordial objects, in particular, the dEs of which quenching of star formation occurred before log ($t_{L}$) $\approx 10$. On the other hand, the origin of some dSphs whose star formation quenched after log ($t_{L}$)  $\approx 9.7$ is not clear. They can be primordial objects as well as transformed ones, If they are primordial objects, the delayed quenching of star formation needs to be explained.

The terminology of fast dwarfs and slow dwarfs were introduced by \citet{gal15} to distinguish cSFHs of dwarf galaxies. On average, the present sample of dSphs and dEs are said to be slow dwarfs because the mean cSFHs at log ($t_{L}$) $\approx 10$ are 0.65 and 0.54 for dSphs and dEs, respectively. However, there are a number of fast dwarfs in the present sample. In particular, a significant fraction of dSphs are fast dwarfs if we consider that dwarfs that quenches at log($t_{L}$) $\approx 10$ or redshift of  $z \sim 2$, are fast dwarfs. Since most star formation occurred before log ($t_{L}$)  $=10$, fast dwarfs are thought to be primordial objects.

There seems to be a trend between the mass of a galaxy and $\tau_{90}$. Less massive dwarfs are likely
to be fast dwarfs \citep{wei14a, wei14b} and they are
considered to be fossils of reionization which formed the bulk of its stars
prior to reionization \citep{wei14b}. The dependence of $\tau_{90}$ on the
mass of a galaxy is consistent with the recent simulations of \citet{dig19}.
The fact that fast dwarfs are more frequent in dSphs than in dEs is consistent with the dependence of quenching time on the stellar mass because dEs are on average $\sim4$ times more massive than dSphs,
Some fast dSphs could be fossils of reionization if the bulk of their stars were formed prior to reionization. The suppression of star formation which resulted in almost complete quenching of star formation in the majority of dSphs after log ($t_{L}$) $\approx 10$ is caused by the feedback from supernova explosions. But, the suppression after the initial bursts in dSphs and dEs at log ($t_{L}$) $>10$ could be caused by reionization or feedback from supernova explosions because both of them depend on the galaxy mass. It is well
known that suppression or quenching of star formation by reionization are
thought to be effective for lower mass dwarfs \citep{gne00, bov09, ben15, jeon17, daw18}. The difference between the SFHs of the two sub-types of early-type dwarfs are thought to be mainly due to their difference in origin.

The pronounced differences in SFHs between dSphs and dEs are the paucity of moderately old ($\sim2.5$ Gyr) stars in dSphs. It is thought to be caused by the effects of reionization and feedback from supernova explosion which expel gas left into the halo of the galaxy. It is also the reason
why the chemical evolution of dSphs was suppressed. These ejected gas evaporated from the halo and some remained there. Since the gas ejected into the halo was enriched before being expelled, this enriched gas can be mixed with the IGM around the galaxy.
The rapid chemical evolution in dEs was possible due to
the relatively large mass which prevented the removal of gas at early epochs
when reionization and supernova feedback were strong.  During the second period
of star formation between log ($t_{L}$) $\approx 9.7$ and $9$, the
metal content of gas out of which the moderately old stars were formed
increased monotonically. The monotonic increase of metallicity in the second
period of star formation can be explained by a simple chemical evolution model
which assumes a closed box with a reservoir such as that of \citet{har76}.
Here, the halo played the role of a reservoir.

\section{Summary and Conclusions}

We have analyzed the SDSS spectra to derive the SFHs of 200 dSphs and 234 dEs of which morphology is classified by \citet{ann15}. On average, dSphs in the present sample is  1.5 mag fainter than dEs.
We used a population synthesis code STARLIGHT \citep{cidF05}.
The SFHs of dSphs and dEs show observational evidences related to the reionization of the Universe
and the stellar feedback. They are pre-enrichment and early quenching of star formation in a significant fraction of early-type dwarfs. The early quenching is more pronounced in dSphs than dEs due to smaller mass of dSphs.  Pre-enrichment of the oldest stellar populations formed at log ($t_{L}$)  $> 10$ made the metallicity of the oldest stellar population as high as Z=0.0004. The non-zero metallicity of the oldest stellar populations is thought to be enriched by PoP III stars during the reionization.

Owing to the energy input from the supernova feedback, there is a period of weakly suppressed star formation in a significant fraction of dSphs and dEs after the initial bursts of star formation at log ($t_{L}$) $> 10$.
The gradual increase of metallicity, from Z = 0.0004 to Z = 0.05, seen in the
stellar populations of dEs which formed until log ($t_{L}$) $\approx 9$ reflects the homogeneous
mixing of metals in the remaining gas of dEs. In contrast,
the star formation in dSphs is almost completely suppressed due to
supernova feedback that removes the gas from the central regions. The gas left after the first period of starbursts is expelled to the halo and some heated gas would be evaporated. It resulted in the paucity of moderately old, intermediate-metallicity stars in dSphs.

The SFH of dEs is characterized by two comparable peaks of active star formation while the SFH of dSphs shows a prominent peak at log ($t_{L}$) $=10$ with a small bump at log ($t_{L}$) $=9.4$. It means that dEs have two periods of active star formation with a gap between them. The reason for the lack of second peak in the SFH of dSphs is the supernova feedback that expels the gas left into the halo of the galaxy or outside the halo. In contrast, dEs can keep a significant amount of gas left for star formation. But it takes time for the gas to cool sufficiently and collapse. This is the reason why there is a gap between the two peaks in the SFH of dEs.

The SFHs of galaxies are thought to be affected by the internal properties of galaxies as well as their environment. 
There is a significant difference in the cSFHs between the early-type dwarfs in the low density regions and those in the high density regions. But if we fix the morphology, the difference between the two density groups decreases significantly although the same trend is maintained. The density dependence of cSFHs of early-type dwarfs is consistent with the dependence of the dynamical times on the local background density. The morphology of early-type dwarfs is thought to be, at least, as important as the stellar mass for the SFHs of the early-type dwarfs. Morphology reflects their SFHs while the stellar mass seems to drive star formation, especially in galaxies with primordial origin. This is the reason why the effect of stellar mass is more pronounced in dSphs which are thought to be mostly primordial objects. 

It is of worth to distinguish dSphs from dEs in morphology classification of early-type dwarf galaxies since their SFHs are clearly different. The difference in SFHs of dSphs and dEs seems to be closely related to their origin, The majority of dSphs are thought to be primordial origin while the majority of dEs are transformed ones.

\section*{Acknowledgments}

Authors would like to thank the anonymous reviewer whose comments and suggestions greatly improve the present paper. This work was supported partially by the NRF Research grant 2015R1D1A1A09057394.  

\section*{Data Availability}
The original data underlying this article are available in SDSS DR7.
We provide the basic output of STARLIGHT as supplemented materials and additional data are available upon request.

% Don't change these lines
\bsp    % typesetting comment
\label{lastpage}

\end{document}